\documentclass[twocolumn,amssymb, amsmath, tightenlines, aps, prl]{revtex4}
\begin{document}
\newcommand{\be}{\begin{equation}}
\newcommand{\ee}{\end{equation}}
\newcommand{\bq}{\begin{eqnarray}}
\newcommand{\eq}{\end{eqnarray}}
\newcommand{\fat}[1]{\mbox{\boldmath $ #1 $\unboldmath}}
\def\qsqrt{{\sqrt{2} \kern-1.2em ^4}}
\def\CC{{\rm\kern.24em \vrule width.04em height1.46ex depth-.07ex
\kern-.30em C}}

\newcommand{\onecolm}{
  \end{multicols}
  \vspace{-3.5ex}
  \noindent\rule{0.5\textwidth}{0.1ex}\rule{0.1ex}{2ex}\hfill
}
\newcommand{\twocolm}{
  \hfill\raisebox{-1.9ex}{\rule{0.1ex}{2ex}}\rule{0.5\textwidth}{0.1ex}
  \vspace{-4ex}
  \begin{multicols}{2}
}

\def\PP{{\rm I\kern-.25em P}}
\def\RR{{\rm
         \vrule width.04em height1.58ex depth-.0ex
         \kern-.04em R}}
\def\id{{\rm 1\kern-.22em l}}
\def\ZZ{{\sf Z\kern-.44em Z}}
\def\NN{{\rm I\kern-.20em N}}
\def\up{\uparrow}
\def\dwn{\downarrow}
\def\L{{\cal L}}
\def\E{{\cal E}}
\def\F{{\cal F}}
\def\ga{{\cal G}[sl(2)]}
\def\H{{\cal H}}
\def\U{{\cal U}}
\def\P{{\cal P}}
\def\M{{\cal M}}
\def\W{{\cal W}}
\def\PACS{\par\leavevmode\hbox {\it PACS:\ }}

\title{Universality of  the one dimensional Bose gas with delta interaction}
\author{Luigi Amico$^{(1)}$ and Vladimir Korepin$^{(2)}$}

\affiliation{$^{(1)}$ MATIS-INFM $\&$  Dipartimento di Metodologie Fisiche e Chimiche (DMFCI), 
        Universit\'a di Catania, viale A. Doria 6, I-95125 Catania, Italy} 
\affiliation{$^{(2)}$ C.N. Yang Institute for Theoretical Physics, State University of New York at Stony Brook,
          NY 11794-3840}  

\begin{abstract}
We consider several models of interacting bosons in a one dimensional lattice.
Some of them are not integrable like  the Bose-Hubbard others are integrable.  
At low density all of these models can be described by the Bose gas
 with delta interaction. The  
lattice corrections corresponding to the different models are contrasted.
\end{abstract}
\maketitle

\section{I. Introduction}

The Bose gas with delta 
interaction is one of the most famous models for strongly interacting 
bosons~\cite{LIEB, YANG} (the corresponding classical dynamics is described 
by the non-linear Schr\"odinger equation). In one dimension this model 
possesses infinitely many, mutually commuting integrals of the 
motion and is therefore  called 
integrable.  The integrability of the Bose gas can be formulated within 
the Quantum Inverse Scattering Method (QISM) \cite{KOREPIN-BOOK} that exploit 
a crucial property of the two-body scattering matrix of the system:
it satisfy the Yang-Baxter equation. This reduces the  
many-body dynamics to a two body dynamics.
In the Appendix we explain the QISM $\&$ the algebraic Bethe Ansatz [it uses a transfer matrix]. 
In the last decades many authors studied integrable  lattice regularization 
of the continuous Bose gas~\cite{IK, KOREPIN-LATTICE,RAGNISCO,Faddeev,KULISH,GERDJIKOV, bb1, bb2, bbp, bbt, bik}. 
The procedure  it is not straightforward since a direct quantization  
of {\it lattice} classical integrable theory whose continuous limit 
is the classical continuous Bose gas field theory leads to ``non-local'' 
Hamiltonians (that is the interaction involves far neighbors). 
This inconvenience  was circumvented in two, somewhat 
complementary ways: keep the scattering matrix of the continuous theory 
unaltered and change 
the transfer matrix and  then  the Hamiltonian \cite{IK, KOREPIN-LATTICE,RAGNISCO,Faddeev}; alternatively,  change the scattering 
matrix keeping the Hamiltonian formally unaltered \cite{KULISH,GERDJIKOV}. In both cases, however, the 
Hamiltonians though (quasi) local,  involve higher order processes whose physics is hard to read-out.

Another paradigmatic lattice model for interacting bosons is  
the Bose-Hubbard  model. It was  proposed first by 
Haldane~ \cite{HALDANE} (see also \cite{FISHER}), and widely used   
in many different contexts.
In mesososcopic physics it describes granular superconductors, Helium films, 
and Josephson junctions arrays~\cite{SARO-REV}.
In quantum optics, the Bose-Hubbard dynamics was proposed to model  
bosonic atoms in  optical 
lattices \cite{OPTICAL-BH} and as effective theory for   higher spin 
chains \cite{DELGADO}.
The Bose-Hubbard Hamiltonian describes charged bosons hopping in a $d$-dimensional
 lattice  and experiencing  a Coulomb repulsion. The resulting  competition between kinetic and electrostatic 
energies induces quantum fluctuations that dominate the  zero temperature phase diagram.   
At {\it commensurate}  number of bosons per site  the Bose-Hubbard model predicts a  
Mott insulator-superfluid  quantum phase transition    
controlled by the ratio between the bosonic tunneling rate $t$ and the  Coulomb repulsion.  
In  optical applications the system parameters are tuned by changing 
the laser intensity \cite{OPTICAL-SI}; the 
criticality of the superfluid-insulator 
quantum phase transition is substantially altered by the confining harmonic potential ~\cite{BATROUNI}. 

In this paper we focus on  one dimension. 
We remark that  $d=1$ is more than an academic case
for the physical systems captured by the models 
under consideration. In fact bosonic chains are commonly 
 realized  both with optical confinement \cite{OPTICAL-SI} and, using a different 
technology, with  Josephson junctions \cite{SARO-REV}. 
Quantum effects are strongest in the one-dimensional case. For example, in the 
dilute limit of optical lattices  quantum many body  effects cause the breakdown of the 
Gross-Pitaevskii theory (valid for  $3d$ systems) and drive the system to the 
so called ``Tonks-Girardeau'' regime \cite{GIRARDEAU} where the exact results 
of Lieb and Liniger are crucial \cite{YNGVASON}. Exact results of the Bose-Hubbard model would be desirable 
to study physical systems like the above  ones, at higher densities.    
However the  exact solution of the one dimensional Bose-Hubbard model is still a challenging problem.
For infinite range hopping the thermodynamics is fully 
established~ \cite{DORLAS}. 
For nearest neighbor hopping,  the coordinate Bethe ansatz 
fails~ \cite{HALDANE-BETHE}  due   to the multi-occupancy characteristic  of the  
bosonic statistics ~\cite{SUTHERLAND}. 
Recent numerical studies suggested  
that the distribution of level spacings obeys  Wigner-Dyson statistics \cite{LEVEL-BOSE}, this  considered as indication  of non-integrability
 \cite{LEVEL-STATISTICS}.

In this paper we discuss the integrable continuous field theory of 
the Bose gas as universal 
low density limit of various integrable 
(Faddeev-Takhtadjan-Tarasov \cite{Faddeev}, 
Izergin-Korepin \cite{KOREPIN-LATTICE}, and quantum 
Ablowitz-Ladik \cite{KULISH,GERDJIKOV,bbp, bbt,bik}) 
and non integrable (Bose-Hubbard) lattice models.
In particular a number of results obtained for the Bose gas  can be used for the behavior of the systems phrased by  Bose-Hubbard models. 
Finally we discuss the differences between the existing {\it integrable}
lattice regularization  of the Bose gas and the non-integrable Bose-Hubbard model. 

The paper is laid out as follow. In the next section we summarize a number of properties of the integrable field theory
of the Bose gas. In section III we apply some of these properties to the Bose-Hubbard model. 
The lattice integrable theories are discussed in section IV. In section V we draw our conclusions. 
In the Appendix we summarize the main ideas of the QISM.

\section{II. The Bose gas with delta interaction}

We shall start with continuous field theory.
This will be the  continuous limit of all bosonic models considered 
in this paper. The one dimensional Bose gas is described by 
an integrable non-relativistic model of  field theory  
 \cite{LIEB,KOREPIN-BOOK}.The Hamiltonian is:  
\begin{equation}
{\cal H}_{BG}=\int dx \left [(\partial_x \Psi^\dagger)(\partial_x \Psi)
+ c \Psi^\dagger\Psi^\dagger\Psi\Psi-h\Psi^\dagger\Psi \right ] \,.
\label{bose-gas}
\end{equation}
Here: $[\Psi (x),\Psi^\dagger (y)]=\delta(x-y)$ and  
$[\Psi^\dagger (x),\Psi^\dagger (y)]=0$.
The coupling constant is denoted by $c$, and $h$ is the chemical potential.
It is equivalent to
a many body  quantum mechanical problem
with  Hamiltonian:
\begin{equation}
H_{BG}=-\sum_{j=1}^N \frac{\partial^2}{\partial z_j^2}+2c\sum_{N\ge j> k \ge 1} \delta (z_j-z_k) \,.
\end{equation}
The energy is $E=\sum_{j=1}^N\left (u_j^2-h \right )$; the momenta $u$'s obey Bethe equations:
\begin{equation}
e^{iu_j L}=\prod_{k\neq j}^N \frac{u_j-u_k+ic}{u_j-u_k-ic}\qquad  j=1\dots N \; .
\end{equation}
Let us consider repulsive case $c>0$. In the thermodynamic limit $N,L\rightarrow \infty$ such that $N/L=D=const$
the  ground state energy-density [per unit length] is 
$$E=\int_{-q}^q\epsilon_0 (u) \rho (u) du =\int_{-q}^q(u^2-h)\rho (u) du\; .$$
The function $\rho(u) $ describes the distribution of
momenta in the ground state. It is defined by the equation:
$$2 \pi \rho(u)=1+ \int_{-q}^q K(u-v) \rho(v) dv \doteq {\cal Z}(u)\;,$$
where ${\cal Z}(u)$ is called dressed charge. 
The kernel of the integral equation is   $K(x)=2c/(c^2+x^2)$. For further purposes it is convenient 
to define a function $\tau(u) $ through 
$$2\pi \tau(u)=K(u-q)-\int_{-q}^q dv K(u-v) \tau (v)\;.$$
The density,
given by $D=\int_{-q}^q \rho(u) du$, depends on the chemical potential 
since $q=q(h)$ \cite{KOREPIN-BOOK}.  
The  Bose gas is characterized by different phases depending on 
the sign of the coupling constant $c$. Let us describe the ground state at zero 
temperature: 

{\it i)} For $c=0$ we have free bosons. In the ground state
all the particles have  zero momentum (Bose-condensation).

{\it ii)} In the repulsive case $ c>0$ the Pauli principle is 
valid (this is the distinctive feature of the one dimensional bosons).
The ground state is a Fermi sphere.

{\it iii)} In the attractive case $c<0$
 the ground state is a single large bound state of all the particles.
All the particles stay close to one another (in configuration space). 
It is a droplet. One can say that in this case the phases separate.

In the repulsive case at zero temperature the 
asymptotics (large distances) of
correlation functions is~ \cite{KOREPIN-BOOK,BOGOLIUBOV-FINITE} 
\begin{eqnarray}
&&\makebox{\hspace{-0.7cm}} \langle \Psi^\dagger (x) \Psi (0) \rangle = A x^{-1/{\theta}} \, \\
&&\makebox{\hspace{-0.7cm}} \langle \Psi^\dagger (x) \Psi (x) \Psi^\dagger (0) \Psi(0) \rangle = D^2+\frac{A}{x^2}+B \frac{\cos{[2\pi D x]}}{x^{\theta}} \, ,
\label{zero-temp-BG}
\end{eqnarray}
where the  critical exponent $\theta$ depends on the density in a remarkably simple way through:
$$\theta=2{\cal Z }^2(q)=4\pi D/v_F\; .$$ 
The thermal correlation functions can be obtained 
by replacing: $x\rightarrow \frac{v_F}{\pi T} \sinh{(\pi T x/v_F)}$ in these 
expression. Here $v_F$ is the Fermi velocity [for the model under 
consideration, it coincides with the velocity of sound]. 
So in the repulsive case  the  asymptotic of the
 thermal correlation functions is
\begin{eqnarray}
&&\langle \Psi^\dagger (x) \Psi (0) \rangle_T = 
\left (\frac{v_F}{\pi T}\right )^{-1/{\theta}} 
\exp{(-\pi T x/v_F{\theta}) } \, ,\\
&&\langle \Psi^\dagger (x) \Psi (x) \Psi^\dagger (0) \Psi(0) \rangle_T = B_2 
\left (\frac{2\pi T}{v_F}\right )^2 e^{-2 \pi T x/v_F}  \nonumber \\ &&\makebox{\hspace{1cm}}  + B_3 \left( \frac{2\pi T}{v_F}\right )^\theta 
e^{-\pi T \theta x/v_F} \cos(2\pi D x) \; . 
\label{temp-BG}
\end{eqnarray}
The coefficients $A,B,B_2,B_3$ are related to certain form factors \cite{BOGOLIUBOV-FINITE}.

\section{III. Application to the Bose-Hubbard models}
   
The Hamiltonian for the one dimensional Bose-Hubbard model reads
\begin{equation}
H_{BH}= \sum_{i=-N_s}^{N_s} [U (n_i-1)-\mu] n_i -t ( a^\dagger_i a_{i+1}+
 a^\dagger_{i+1} a_{i}) \,,
\label{bose-hubbard}
\end{equation}
where the operators $n_{i} := a^{\dagger}_{i} a_{i}$ count the
number of bosons at the site $i$; operators $a_{i}$, $a^{\dagger}_{i}$ obey
the canonical commutation relations $[a_{i},a^{\dagger}_{j}]= \delta_{ij}$ and 
and $2N_s+1$ is the number of sites.
The parameters $t$, $U$ of (\ref{bose-hubbard}) are the hopping
amplitude and the strength of the on-site Coulomb repulsion,
respectively, while the chemical potential $\mu$ 
fixes the average number of bosons in each site.
The density of bosons in the lattice is given by $D=N/(2N_s+1) \Delta$. Here
$\Delta$ is the lattice  spacing.
The Bose-Hubbard model can be considered as a  possible 
lattice regularization of the Bose gas. 
In fact at  small filling factor $\nu=N/(2N_s+1)=D \Delta$ the Bose-Hubbard model 
can be described  by the Bose gas  with delta interaction.
In order to obtain the correct commutators, in the continuous limit $\Delta \rightarrow 0$
one has to re-normalize the operators: $a_i=\sqrt \Delta \Psi (x)$
and $n_i= \Delta \Psi^\dagger (x)\Psi (x)  $,  $x=\Delta i$.
The Bose-Hubbard model  then reduces to the Bose gas
$$H_{BH}=t\Delta^2 H_{BG}\;,$$
with  $c=U/(t\Delta)$ and $h=-(\mu +2 t)/(t\Delta^2)$. Therefore results for 
the Bose gas  can be phrased  for systems captured by the Bose-Hubbard model.
The  asymptotics of the  zero-temperature correlations in the repulsive case are
\begin{eqnarray}
\hspace*{-0.5cm}\langle a_i^\dagger a_j \rangle &=& A (i-j)^{-{1}/{\theta}} \,,\\
\hspace*{-0.5cm} \langle n_i n_j \rangle &= &(\frac{\nu}{\Delta})^2+\frac{A}{(i-j)^2}+B \frac{\cos{[2\pi \nu (i-j)/\Delta]} }{(i-j)^{\theta}}\, ,
\end{eqnarray} 
where the  critical exponent $\theta$ is
\begin{equation}
\theta={2}\left (1+\frac{4D}{ c}\right ) = {2}\left (1+\frac{4\nu t}{ U}\right ) \; .
\end{equation}
The thermal   correlation functions  read
\begin{eqnarray}
&&\makebox{\hspace{-0.6cm}}\langle a_i^\dagger a_j  \rangle_T = \left (\frac{v_F}{\pi T}\right )^{-1/{\theta}} 
\sinh{\left [\pi T (i-j)/v_F\right ]}^{-1/{\theta}} \,,\\
&&\makebox{\hspace{-0.6cm}}\langle n_i n_j \rangle_T = B_2 
\left (\frac{2\pi T}{v_F}\right )^2 e^{-2 \pi T (i-j)/v_F}  \\ &&\makebox{\hspace{0.2cm}}  
+ B_3 \left(\frac{2\pi T}{v_F}\right )^\theta 
e^{-\pi T \theta (i-j)/v_F} \cos\left [2\pi \nu  (i-j)/\Delta\right ] \; . \nonumber
\label{temp-BH}
\end{eqnarray}

\section{IV. Integrable lattice models}
In this section we discuss the integrable bosonic theories  arising as quantum lattice regularization of the field theory of the Bose gas 
with delta interaction. From a classical side the Hamiltonian 
 structure [action-angle variables]
is directly related  to the $R$-matrix. This constituted motivation to define the lattice models keeping unaltered 
the $R$-matrix (for an alternative way to construct lattice regularization of the Bose gas see the Appendix~\cite{KULISH,GERDJIKOV}). 
Pursuing a  quantum version of this procedure, namely  directly quantizing 
the classical Lax matrices, it turns out that the arising Hamiltonians  
are  non-local.
Here we should note that  non-local  Hamiltonians are interesting by them-self:
good examples are Calogero-Sutherland, Haldane-Shastry~\cite{POLY}, and 
pairing~\cite{RICHARDSON,AMICO,SIERRA} models.
To obtain local Hamiltonians the Lax matrices have to be modified. 
This modification induces complications on the final form of the though 
local Hamiltonians (see the Refs.~\cite{KOREPIN-LATTICE,IK,RAGNISCO,Faddeev}). 
All these  are solvable 
by  algebraic Bethe Ansatz (see Ref.~ \cite{KOREPIN-BOOK}); 
some basic formulas of the QISM  are summarized in Appendix.
Here we will present  these  Hamiltonians. 
Some of them simplify in the weak coupling limit. In this limit the resulting 
Hamiltonians are compared with the Bose-Hubbard model.

\subsection{Izergin-Korepin models}  

The first integrable version of  the Bose gas on the lattice 
was constructed in Refs~ \cite{IK, KOREPIN-LATTICE}. 
To write the Hamiltonian of the model in a compact way we introduce 
the following notations. The bosons interact 
differently in odd and even lattice sites. Let us start with canonical Bose
 operators
\begin{equation}
[a_j,a^\dagger_k]=\delta^j_k \quad ; \qquad a_j|0\rangle=0 \; .
\end{equation}
In order to describe the model, it is convenient to introduce 'renormalized'
operators 
\begin{equation}
b_j=a_j\sigma^{-1}_j\; .
\end{equation}
 Here $\sigma_j$ is slightly different in odd and even lattice sites:
\begin{equation}
\sigma_j=\left\{ \begin{array}{ll} 
\sqrt{1+\omega a^\dagger_ja_j} & \mbox{if $j$ is even} \, ,\\
\sqrt{1+\omega (a^\dagger_ja_j-1)} & \mbox{if $j$ is odd}\,  .
\end{array}
\right.
\end{equation}
where  $\omega= c\Delta/4$.
We can express everything is terms of 'renormalized' operators $b_j$:
\begin{eqnarray}
a^\dagger_ja_j & =& b^\dagger_jb_j \{ 1-\omega  b^\dagger_jb_j \}^{-1} \,  ,\\
\sigma^2_j & =& \{ 1-\omega  b^\dagger_jb_j \}^{-1} \quad \mbox{for $j$ even}\;,
\end{eqnarray}
For odd $j$ the expressions are slightly  different:
\begin{eqnarray}
a^\dagger_ja_j & =&(1-\omega) b^\dagger_jb_j \{ 1-\omega  b^\dagger_jb_j \}^{-1}\,,\\
\sigma^2_j & =&(1-\omega) \{ 1-\omega  b^\dagger_jb_j \}^{-1} \quad \mbox{for $j$ odd}\;.
\end{eqnarray}
We can also write a closed commutation relations for renormalized operators:
\begin{eqnarray}
b_j  \{ 1&-&\omega  b^\dagger_jb_j \}^{-1}b^\dagger_j - \{ 1-\omega  b^\dagger_jb_j \}^{-{1\over 2}}  b^\dagger_jb_j \{ 1-\omega  b^\dagger_jb_j \}^{-{1\over 2}} \nonumber \\
&=& \left\{ \begin{array}{ll}
1 \quad & \mbox{if $j$ is  even }\\
(1-\omega)^{-1} \quad & \mbox{if $j$ is odd } \; .
\end{array}
\right.
\end{eqnarray}
In different lattice sites operators $b_j$ commute.
Now we can present the Hamiltonian of the model as
\begin{equation}
H_{IK}=-\frac{4} {3c\Delta^3} \sum_j \left( t_j + t_j^{\dagger} +{2-\omega \over 2(1-\omega)} -{\Delta  a^\dagger_ja_j\over 1-\omega^2}\right) \,.
\end{equation}
Below, we shall represent operators $t_j$ in the form
\begin{equation}
t_j=-{1\over 2} O_j^{-1} C_j  O_j \;. 
\end{equation}
The formula for operators $O$ and $C$ are different for odd and even lattice 
sites. For $j$ odd:
\begin{eqnarray}
O_j&=&\sigma_{j+1} \{ 1+\omega b^\dagger_{j+1} b_{j+2} \}\sigma_{j+2} \,, \\
C_{j}&=&\sigma^{-1}_j\{ 1+\omega  b^\dagger_{j-1} b_{j}\}^{-1}\sigma^{-2}_{j-1}
\{ 1+\omega  b^\dagger_{j} b_{j-1}\}^{-1} \times \nonumber \\
& &\sigma^{-1}_j \sigma_{j-1} \{ 1-\omega  b^\dagger_{j-1} b_{j+1}\}  \sigma_{j+1} \,,.
\end{eqnarray}
For $j$ even:
\begin{eqnarray}
O_j&=&\sigma_{j-2} \{ 1+\omega b^\dagger_{j-2} b_{j-1} \}\sigma_{j-1} \,, \\
C_{j}&=&\sigma^{-1}_{j-1}\{ 1+\omega  b^\dagger_{j} b_{j-1}\}^{-1}\sigma^{-2}_{j}
\{ 1+\omega  b^\dagger_{j-1} b_{j}\}^{-1}  \times \nonumber \\
& & \{ 1-\omega  b^\dagger_{j-1} b_{j+1}\}  \sigma_{j+1} \,.
\end{eqnarray}
In the weak coupling limit the Hamiltonian describes a  coupling of
five neighbors lattice sites $j-2 \dots j+2$. 
Retaining  contributions up to second order in $c\Delta$ it  has 
the form 
\begin{eqnarray}\label{IK-weak}
H_{IK}&=& -\frac{4} {3c\Delta^3}   \sum_j \frac{c\Delta}{8} 
(K_{j,j-1}-n_j-K_{j-1,j+1} ) \\
&&\hspace{3cm}+\frac{(c\Delta)^2}{16} 
( K_{j,j-1}^2+ {h}_j) \,,\nonumber \\
{h}_j&=&\sum_{\alpha,\beta=-2}^2 \left [
v_{\alpha \beta} n_{j+\alpha} n_{j+\beta}+ w_{\alpha \beta} 
a_{j+\alpha} a^\dagger_{j+\beta}\right .\\
&&\hspace{-2cm} \left. +\left (t_{\alpha}+
q_{\alpha} n_{j+\alpha} \right )
\left (r_{\beta} a_{j+\beta} a^\dagger_{j+\beta-1} +
s_{\beta}a_{j-\beta} a^\dagger_{j+\beta}\right )\right ] +h.c. \; ,\nonumber
\end{eqnarray}
where $K_{l,m}=a_l a^\dagger_{m} +a_{m} a^\dagger_{l}$
(with certain coefficients $v_{\alpha \beta}\,,  w_{\alpha \beta} \, , \:t_{\alpha \beta}\, , \; q_{\alpha \beta}$). 
After simple manipulations Eq.~(\ref{IK-weak}) can be written as
\begin{eqnarray}
H_{IK}=H_{BH}-g \frac{c\Delta}{8} K_{j-1,j+1}+ g (\frac{c\Delta}{4})^2 
\left [ K_{j,j-1}^2+ h_j \right ]\,,\nonumber \\
\phantom{}
\end{eqnarray}
where periodic boundary conditions have been considered: 
\begin{equation}
\sum_j n_j=\sum_j n_{j+1} \quad \makebox{and} \quad \sum_j n_j^2=\sum_j n_{j+1}^2\;;
\label{PBC}
\end{equation}
the parameters  in  $H_{BH}$  above are $\mu/g=(c \Delta)  
(1- 5c \Delta /16)/8$,  
$U/g= 5 ( c \Delta)^2/128 $, $t/g=c\Delta/8$ with $g=-4/(3c\Delta^3)$. 
The Bose-Hubbard model differs from Itzergin-Korepin model 
in the presence of the quadratic hopping and in  the non local terms.  
In the continuous limit, when the lattice spacing $\Delta\rightarrow 0$, 
this lattice model turns into a continuous Bose gas (\ref{bose-gas}).

\subsection{ Faddeev-Takhtadjan-Tarasov models}

In this section we discuss  another lattice integrable version
of the Bose gas. The Hamiltonian was suggested by
Faddeev-Takhtadjan-Tarasov in Ref.~ \cite{Faddeev}.
The model has the same $R$-matrix as the continuous Bose gas.
 It looks like a quantum  spin chain with
negative spin. Let us first introduce an operator of 'angular-momentum'
$J_{j,j+1}$. It is defined as a solution of the following operator equation:
$$J_{j,j+1}(J_{j,j+1}+1)=2S_j\otimes S_{j+1}+2s(s+1).$$
Here the 'spin' is $s=-2/(c\Delta )$.
The Hamiltonian of the  Faddeev-Takhtadjan-Tarasov model is 
$$H_{FTT}=-2 \kappa \sum_j {\Gamma'(J_{j,j+1}+1)}/{\Gamma(J_{j,j+1}+1)}\; .$$ 
The relation with the lattice bosons becomes  transparent using the 
Holstein-Primakov realization of spins:
\begin{eqnarray}
S^x_j&=& { a^\dagger_j \rho_j+\rho_j a_j \over \sqrt{c \Delta}}\,, \nonumber \\
S^y_j&=&i{-a^\dagger_j \rho_j+\rho_j a_j\over \sqrt{c \Delta}}\,, \\
S^z_j&=&-{2\over c\Delta}\left( 1+{c \Delta a^\dagger_j a_j\over 2}\right) \,,\nonumber 
\end{eqnarray}
with $\rho=\sqrt {1+c \Delta a^\dagger a /4}$. 
 For small $c $, the  spin $s$ is large.
The Hamiltonian simplifies (second order in $1/s$):
\begin{eqnarray}
\label{FTT-hamiltonian}
\hspace{-3cm}&H_{FTT}=\displaystyle{\frac{\kappa}{s^3}}\sum_j \left \{ {s^2\over 4}(a^\dagger_{j+1}-a^\dagger_{j})(a_{j+1}-a_{j})
\right. \nonumber \\ 
 & \hspace{-0.2cm}\left. -\displaystyle{{s\over 16}} [ (a^\dagger_{j})^2 + a^\dagger_{j+1}a^\dagger_{j}+
(a^\dagger_{j+1})^2] [ a_{j}^2 + a_{j+1}a_{j}+ a_{j+1}^2 ]\right. \nonumber \\
&\left.+\displaystyle{{s\over 16}}a^\dagger_{j+1} a^\dagger_{j}a_{j+1} a_{j} -s^2 a^\dagger_{j}a_{j}\right \} \;.
\end{eqnarray}
The model is solvable by algebraic Bethe Ansatz \cite{Faddeev}.
After simple manipulations Eq.(\ref{FTT-hamiltonian}) can be written as
\begin{eqnarray}
\label{FTT-BH}
&&H_{FTT}=H_{BH} \\ 
&&+ V  \sum_j \left [n_j n_{j+1} +\frac{1}{2}K_{j,j+1}^2 
+ (n_j+n_{j+1} )K_{j,j+1} \right ] \,,\nonumber
\end{eqnarray}
where periodic  boundary conditions have been considered (see Eqs.(\ref{PBC}));
the parameters  in  $H_{BH}$  above are $\mu=\kappa (c \Delta)  ( c \Delta /16+1)/4$,  
$U=\kappa ( c \Delta)^2/64 $, $t=\kappa c\Delta/8$, and $V=\kappa {(c \Delta)^2}/{64}$.
Eq.~(\ref{FTT-BH}) elucidates the relation between the Faddeev-Takhtadjan-Tarasov and Bose-Hubbard models.
In the continuous limit, when lattice spacing $\Delta\rightarrow 0$ 
the Faddeev-Takhtadjan-Tarasov  (and the Bose-Hubbard model as well) lattice model turns into the  
continuous Bose gas (\ref{bose-gas}).

\section{V. Conclusions.} 
We considered several models of interacting bosons.  
Faddeev-Takhtadjan-Tarasov model and Izergin-Korepin model are different lattice models originally  constructed from 
the Bose gas continuous field theory. For these models the lattice 
contributions do not destroy the 
integrability of the original field theory; instead the Bose Hubbard model contains lattice-effects leading to a non 
integrable dynamics. In the weak coupling limit these integrable Hamiltonians 
and the Bose-Hubbard Hamiltonian can be  compared  
explicitly. 
In particular the model in Eq.(\ref{FTT-BH}) could serve  exact studies 
of chains of Josephson junctions 
away from the degeneracy point \cite{SARO-REV}.
The lattice effects vanish at low density: all these theories are integrable and described 
by the Bose gas with delta interaction. 
In this regime the known asymptotic of the Bose gas correlation functions can be used for physical systems captured by the 
Bose-Hubbard model (see~ \cite{ZWERGER} for a recent application). The Luttinger liquid field theory in 
Ref.~ \cite{HALDANE-1981} and the Bose gas are characterized 
by the same set of correlation functions where the spectral parameters $v_J,v_F,v_N$ are determined by \cite{HALDANE-BAvsLL}
$\theta=2\sqrt{v_J/v_N}$, $v_F=\sqrt{v_N v_J}$, and   
$v_F=\left [ \epsilon_0'(q)-\int_{-q}^q  du \epsilon_0'(u)\tau(u)\right ]/[{\cal Z}(q)]$ (see section II).

\bigskip
Discussions with R. Fazio, P. Kulish  and A. Osterloh  are acknowledged.

\section{Appendix}

The construction of integrable lattice models of interacting bosons
is  based on  Quantum Inverse Scattering Method (QISM)~ \cite{KOREPIN-BOOK}.
Here we shall summarize some features of this method.
The starting point of the QISM is  a local quantum Lax operators
$L_i(u)$ and  a matrix $R(u)$, satisfying the Yang 
Baxter  equation
$$
R(\phi) L_i(\zeta)\otimes L_i(\xi)=L_i(\xi)\otimes L_i(\zeta) 
R(\phi)\; .
$$
The monodromy matrix 
\begin{equation}
T(\zeta)=L_{-N_s}(\zeta) L_{-N_s+1}(\zeta)  \dots L_{N_s}(\zeta)\;, 
\label{monodromy}
\end{equation} 
also fulfills the Yang-Baxter relation  
\begin{equation}
R(\phi) T(\zeta)\otimes T(\xi)=T(\xi)\otimes T(\zeta) 
R(\phi)\;.
\end{equation}
The transfer matrix is defined as
$
t (\zeta):=tr_{(0)} T(\zeta) \;  
$
where $tr_{(0)}$ means the trace  in the auxiliary 
space. It is a generating functional of integrals 
of motion and of the Hamiltonian since
it commutes with itself  at different values of spectral parameters:
$[t (\zeta),t (\xi)]=0$ ($t (\zeta)$ is an invariant of the Yang-Baxter 
algebra). 
The quantum determinant~ \cite{QUANTUM-DET} 
\begin{equation}
{\det}_{q}(T(\zeta):={\rm Tr} {\cal P }T(\zeta)\otimes T(e^{-\eta}\zeta) \,,
\label{q-det}
\end{equation}
is the analog of the Casimir operator of  Lie algebras and it is   
another   invariant of the Yang-Baxter algebra. In Eq.~(\ref{q-det})  
${\cal P}$ is the projector 
\begin{equation}
{\cal P}=R(\eta;\eta):= {1\over 2\cosh \eta} \left(
\begin{array}{cccc}
0 &  0 & 0 & 0   \\
0 &  e^{-\eta} & -1 & 0  \\
0 & -1 & e^\eta &0 \\
0&  0 & 0   &0
\end{array}
\right) \,. 
\label{projector}
\end{equation}

To construct lattice integrable theories for the Bose gas two 
different procedures (somehow complementary) have been pursued. 
The first consists in keeping the $R$-matrix of the continuous theory 
unaltered and changing the Lax operators. Such procedure has been followed 
in the text above.
One can also change  both the $R$ and $L$ operators to simplify the final
 form of the Hamiltonian.
This approach was initiated by P.P. Kulish {\it et al.} \cite{KULISH,GERDJIKOV} and 
developed in the papers \cite{bbp,bbt,bik}, quantizing classical discretization on non-linear Schroedinger 
equation developed earlier by Ablowitz and Ladik \cite{ABLOWITZ}.
For
the Ablowitz-Ladik   model the Lax operators and the  R- matrix \cite{KULISH,GERDJIKOV} are :
\begin{equation}
L_i(\zeta):= \left(
\begin{array}{cc}
\zeta &  q_i   \\
-q_i^\dagger  &\zeta^{-1}   
\end{array}
\right)  \,,
\
R(\phi;\eta):= \left(
\begin{array}{cccc}
1 &  0 & 0 & 0   \\
0 &  b^- & c & 0  \\
0 & c & b^+ &0 \\
0&  0 & 0   &1
\end{array}
\right) \,.
\label{Ablowitz}
\end{equation}
Here $\exp\phi={\zeta}/\xi$ and  $\zeta, \xi $ are  spectral parameters,
$$b^\pm={e^{\pm \eta} \sinh{\phi}\over \sinh{(\phi-\eta)}}\,,$$ 
$$c=-{\sinh{\eta}\over \sinh{(\phi-\eta)}}\,,$$ the quantity $\eta\in \RR$ is a
deformation 
parameter. The Hamiltonian of quantum Ablowitz-Ladik can be defined by 
\begin{equation}
H_{AL}=-\sum_{j=-N_s}^{N_s} \left [q^\dagger_j q_{j+1}+q^\dagger_{j+1} q_j -\alpha  
\log (1+q^\dagger_j q_j)\right ]\label{q-AL} \,.
\end{equation}
We shall see later that $\alpha$ doesn't play the role of  coupling
 constant. In fact the potential and the kinetic energies commute
(then the potential energy turns into a constant on the eigenstates
of the kinetic term of (\ref{q-AL})).
The operators $q_j$ satisfy the following commutation relation:
\begin{equation}
[q_i,q^\dagger_j]=(e^{2\eta}-1) \left (1+q^\dagger_j q_j \right ) \delta^i_j 
\end{equation}
as follows the Yang-Baxter algebra. The Hamiltonian~(\ref{q-AL}) 
can be rewritten using the trace identities. In particular the 
hopping term of the Ablowitz-Ladik  model is  
\begin{equation} 
-\sum_j q_j q^\dagger_{j+1} = \lim_{\zeta \rightarrow \infty}\frac{t(\zeta)-\zeta^{2N_s+1}}{\zeta^{2 N_s-1}} \;,  
\end{equation}
and
\begin{equation} 
-\sum_j q^\dagger_j q_{j+1} = \lim_{\zeta\rightarrow 0} \frac{t(\zeta)-
\zeta^{-2N_s-1}}{\zeta^{1-2 N_s}} \,,
\end{equation}
whereas the interaction is related to the quantum determinant
\begin{equation}
\ln{\det}_{q}(T)=\sum_{i=-N_s}^{N_s}\log (1+q^\dagger_j q_j)+(2N_s+1)\eta \label{qd}\,.
\end{equation}
In simple words:  the hopping term (the kinetic energy) commutes with the interaction since 
the transfer 
matrix and the quantum determinant commute each other.
Eigenvectors of the  transfer matrix can be constructed by algebraic Bethe
Ansatz (see Refs.~ \cite{KULISH,bbp} or Chapter VII of the book \cite{KOREPIN-BOOK}).
Let us denote the off-diagonal element of monodromy matrix (\ref{monodromy})
by $C(\zeta)=T_{21}(\zeta) $. The eigenvectors of monodromy matrix and Hamiltonian can be written as
\begin{equation} 
|\chi\rangle =\prod_{k=1}^n C(\zeta_k)|0\rangle \label{ev}\,.
\end{equation}
Here $|0> $ is  the Fock vacuum $q_j|0\rangle =0$.
The eigenvalue of monodromy matrix $\theta (\zeta)$ is
\begin{eqnarray} 
\theta (\zeta)=& \zeta^{2N_s+1}e^{n\eta}\prod_{k=1}^n{\sinh (\lambda -\lambda_k-\eta)\over \sinh (\lambda -\lambda_k) }\nonumber \\
 + & \zeta^{-2N_s-1}e^{n\eta}\prod_{k=1}^n{\sinh (\lambda -\lambda_k+\eta)\over \sinh (\lambda -\lambda_k) } \,.
\end{eqnarray}
Here  $\exp(\lambda)=\zeta$ (the same for  $\exp(\lambda_k)=\zeta_k$).
The variables  $\lambda_k$ have to satisfy the Bethe equations
\begin{equation}
e^{2(2N_s+1)\lambda_a}=-\prod_{k=1}^n{\sinh(\lambda_a-\lambda_k +\eta) \over
\sinh(\lambda_a-\lambda_k -\eta)} \,.
\end{equation}
 The vectors (\ref{ev}) are also eigenvectors of
the logarithm of the quantum determinant (\ref{qd}) with eigenvalues equal
to $(2N_S+1)\eta$. The energy levels [ eigenvalues of the Hamiltonian]
are
\begin{equation}
2(1-e^{2\eta})\sum_{k=1}^n\cosh 2\lambda_k \,.
\end{equation}
The operators 
$q_j$ can be represented in terms of the standard  Bose operators $[a_n,a^\dagger_m]=\delta^n_m $
$$q={\sqrt \frac{\exp{2 \eta (a^\dagger a +1)}-1}{1+a^\dagger a }}\quad a \; .$$
The correlation functions of the model were evaluated in \cite{bbt,bik}.

\end{document}